\RequirePackage{rotating}
\documentclass[fleqn,usenatbib]{mnras}

\usepackage[T1]{fontenc}
\usepackage{afterpage}
\DeclareRobustCommand{\VAN}[3]{##2}
\let\VANthebibliography\thebibliography
\def\thebibliography{\DeclareRobustCommand{\VAN}[3]{##3}\VANthebibliography}

\usepackage{graphicx}	
\usepackage{amsmath}	
\usepackage{amssymb}	

\usepackage{graphicx}
\usepackage{lineno,hyperref}
\usepackage{float}
\usepackage{color}
\usepackage{textcomp}
\usepackage{nicefrac}
\usepackage{makecell}
\modulolinenumbers[5]
\usepackage{enumitem}
\usepackage{afterpage}
\usepackage{rotating}
\usepackage{pdflscape}

\setlist{noitemsep} 

\title[$^{21}$Ne levels and the enhanced $s$-process]{The impact of $^{17}$O + $\alpha$ reaction rate uncertainties on the s-process in rotating massive stars}

\author[J. Frost-Schenk et al.]{
J. Frost-Schenk,$^1$,
P. Adsley,$^{2,3,4,5}$
A.M.~Laird,$^1$
R.~Longland,$^{6,7}$
C.~Angus,$^{1,8}$
C.~Barton,$^1$
A.~Choplin,$^9$\newauthor
C. Aa.~Diget,$^1$
R. Hirschi,$^{10,11}$
C.~Marshall,$^{6,7}$
F.~Portillo Chaves,$^{6,7}$
K.~Setoodehnia,$^{6,7}$
\\
$^1$Department of Physics, University of York, York, YO10 5DD, UK,\\
$^2$School of Physics, University of the Witwatersrand, Johannesburg 2050, South Africa,\\
$^3$iThemba LABS, National Research Foundation, PO Box 722, Somerset West 7129, South Africa\\
$^4$Cyclotron Institute, Texas A\&M University, College Station, Texas 77843, USA\\
$^5$Department of Physics and Astronomy, Texas A\&M University, College Station, Texas 77843, USA\\
$^6$Department of Physics North Carolina State University, Raleigh, North Carolina 27695-8202, USA\\
$^7$Triangle Universities Nuclear Laboratory, Durham, North Carolina, 27708-0308, USA\\
$^8$TRIUMF, 4004 Wesbrook Mall, Vancouver, BC V6T 2A3\\
$^9$Institut d'Astronomie et d'Astrophysique, Universit\'e Libre de Bruxelles,  CP 226, B-1050 Brussels, Belgium\\
$^{10}$Astrophysics Group, Lennard-Jones Laboratories, Keele University, Keele ST5 5BG, UK\\
$^{11}$Kavli IPMU (WPI), The University of Tokyo, Kashiwa, Chiba 277-8583, Japan\\
}

\begin{document}
\label{firstpage}
\pagerange{\pageref{firstpage}--\pageref{lastpage}}
\maketitle

\let\jnl=\rm
\def\reff@jnl#1{{\jnl#1}}

\def\aj{\reff@jnl{AJ}}                   
\def\actaa{\reff@jnl{Acta Astron.}}      
\def\araa{\reff@jnl{ARA\&A}}             
\def\apj{\reff@jnl{ApJ}}                 
\def\apjl{\reff@jnl{ApJ}}                
\def\apjs{\reff@jnl{ApJS}}               
\def\ao{\reff@jnl{Appl.~Opt.}}           
\def\apss{\reff@jnl{Ap\&SS}}             
\def\aap{\reff@jnl{A\&A}}                
\def\aapr{\reff@jnl{A\&A~Rev.}}          
\def\aaps{\reff@jnl{A\&AS}}              
\def\azh{\reff@jnl{AZh}}                 
\def\baas{\reff@jnl{BAAS}}               
\def\bac{\reff@jnl{Bull. astr. Inst. Czechosl.}}
\def\caa{\reff@jnl{Chinese Astron. Astrophys.}}
\def\cjaa{\reff@jnl{Chinese J. Astron. Astrophys.}}
\def\icarus{\reff@jnl{Icarus}}           
\def\jcap{\reff@jnl{J. Cosmology Astropart. Phys.}}
\def\jrasc{\reff@jnl{JRASC}}             
\def\memras{\reff@jnl{MmRAS}}            
\def\mnras{\reff@jnl{MNRAS}}             
\def\na{\reff@jnl{New A}}                
\def\nar{\reff@jnl{New A Rev.}}          
\def\pra{\reff@jnl{Phys.~Rev.~A}}        
\def\prb{\reff@jnl{Phys.~Rev.~B}}        
\def\prc{\reff@jnl{Phys.~Rev.~C}}        
\def\prd{\reff@jnl{Phys.~Rev.~D}}        
\def\pre{\reff@jnl{Phys.~Rev.~E}}        
\def\prl{\reff@jnl{Phys.~Rev.~Lett.}}    
\def\pasa{\reff@jnl{PASA}}               
\def\pasp{\reff@jnl{PASP}}               
\def\pasj{\reff@jnl{PASJ}}               
\def\rmxaa{\reff@jnl{Rev. Mexicana Astron. Astrofis.}}%
\def\qjras{\reff@jnl{QJRAS}}             
\def\skytel{\reff@jnl{S\&T}}             
\def\solphys{\reff@jnl{Sol.~Phys.}}      
\def\sovast{\reff@jnl{Soviet~Ast.}}      
\def\ssr{\reff@jnl{Space~Sci.~Rev.}}     
\def\zap{\reff@jnl{ZAp}}                 
\def\nat{\reff@jnl{Nature}}              
\def\iaucirc{\reff@jnl{IAU~Circ.}}       
\def\aplett{\reff@jnl{Astrophys.~Lett.}} 
\def\apspr{\reff@jnl{Astrophys.~Space~Phys.~Res.}}
\def\bain{\reff@jnl{Bull.~Astron.~Inst.~Netherlands}} 
\def\fcp{\reff@jnl{Fund.~Cosmic~Phys.}}  
\def\gca{\reff@jnl{Geochim.~Cosmochim.~Acta}}   
\def\grl{\reff@jnl{Geophys.~Res.~Lett.}} 
\def\jcp{\reff@jnl{J.~Chem.~Phys.}}      
\def\jgr{\reff@jnl{J.~Geophys.~Res.}}    
\def\jqsrt{\reff@jnl{J.~Quant.~Spec.~Radiat.~Transf.}}
\def\memsai{\reff@jnl{Mem.~Soc.~Astron.~Italiana}}
\def\nphysa{\reff@jnl{Nucl.~Phys.~A}}   
\def\physrep{\reff@jnl{Phys.~Rep.}}   
\def\physscr{\reff@jnl{Phys.~Scr}}   
\def\planss{\reff@jnl{Planet.~Space~Sci.}}   
\def\procspie{\reff@jnl{Proc.~SPIE}}   

\let\astap=\aap
\let\apjlett=\apjl
\let\apjsupp=\apjs
\let\applopt=\ao

\begin{abstract}
Massive stars are crucial to galactic chemical evolution for elements heavier than iron. Their contribution at early times in the evolution of the Universe, however, is unclear due to poorly constrained nuclear reaction rates. The competing $^{17}$O($\alpha,\gamma$)$^{21}$Ne and $^{17}$O($\alpha,n$)$^{20}$Ne reactions strongly impact weak s-process yields from rotating massive stars at low metallicities. 
Abundant $^{16}$O absorbs neutrons, removing flux from the s-process, and producing $^{17}$O. The $^{17}$O($\alpha,n$)$^{20}$Ne reaction releases neutrons, allowing continued s-process nucleosynthesis, if the $^{17}$O($\alpha,\gamma$)$^{21}$Ne reaction is sufficiently weak.
While published rates are available, they are based on limited indirect experimental data for the relevant temperatures and, more importantly, no uncertainties are provided.
The available nuclear physics has been evaluated, and combined with data from a new study of astrophysically relevant $^{21}$Ne states using the $^{20}$Ne($d,p$)$^{21}$Ne reaction.
Constraints are placed on the ratio of the ($\alpha,n$)/($\alpha,\gamma$) reaction rates with uncertainties on the rates provided for the first time.
The new rates favour the ($\alpha,n$) reaction and suggest that the weak s-process in rotating low-metallicity stars is likely to continue up to barium and, within the computed uncertainties, even to lead.
\end{abstract}




\section{Introduction}
\label{sect:intro}

Massive stars are key contributors to the abundance of chemical elements, producing elements up to the iron group via charged-particle reactions during their evolution and subsequent explosion in core-collapse supernovae, and synthesising elements heavier than iron via neutron-capture reactions. The weak s-process during core-helium \citep{Langer89, Prantzos90, Baraffe92} and to a smaller extent during shell-carbon burning \citep{Raiteri91b, The07} and possibly the weak r-process in the supernova explosion \citep{2011PrPNP..66..346T} produce elements up to the strontium peak in standard, non-rotating, models. 

Rotation in massive star models significantly boosts the efficiency of the weak s-process, especially at low metallicity \citep{Pignatari08}, enhancing production of elements above strontium. 
With rotation, the helium core contribution to the s-process was shown to increase at the expense of the carbon-burning shell. In the models of \cite{Frischknecht16}, the carbon-burning shell contribution is less than 10~\% 
at sub-solar metallicities.
During the core helium burning phase, rotation-induced mixing transports $^{12}$C and $^{16}$O from the He-core to the H-shell, leading to substantial overproduction of $^{13}$C and $^{14}$N through the CNO cycle. These isotopes are later engulfed by the growing convective He-burning core, leading to a more efficient activation of neutron sources through the $^{14}$N($\alpha,\gamma$)$^{18}$F($\beta+$)$^{18}$O($\alpha,\gamma$)$^{22}$Ne($\alpha,n$)$^{25}$Mg chain and the $^{13}$C($\alpha,n$) reaction \citep{Pignatari08, Frischknecht16, Choplin18, Limongi18, Banerjee19}. 
Specifically, \cite{Cescutti13} have shown that the Sr/Ba scatter at low metallicity of the observed Milky Way halo stars can be reproduced if the contribution of fast rotating massive stars is included. 

The contributions of rotating massive stars to chemical evolution are subject to stellar and nuclear uncertainties. One key nuclear uncertainty is the ratio between the $^{17}$O($\alpha,n$)$^{20}$Ne and $^{17}$O($\alpha,\gamma$)$^{21}$Ne reaction rates \citep{Frischknecht12, Choplin18}. 
At all metallicities, $^{16}$O is abundant in both the helium-burning core and the shell carbon burning.
It is a strong neutron absorber, producing copious amounts of $^{17}$O and loss of neutron flux. The neutrons absorbed by $^{16}$O may be recovered via the $^{17}$O($\alpha,n$)$^{20}$Ne reaction. Alternatively, the $^{17}$O($\alpha,\gamma$)$^{21}$Ne reaction permanently absorbs the neutron, preventing it from contributing to s-process nucleosynthesis. The
$^{17}$O($\alpha,n$)$^{20}$Ne to $^{17}$O($\alpha,\gamma$)$^{21}$Ne ratio, therefore, determines the fraction of neutrons released and the strength of the s-process. 

At He-core burning temperatures (0.2-0.3 GK), the $^{17}$O+$\alpha$ reactions are dominated by resonant contributions from states between $E_x = 7600$ and $8100$ keV ($E_r \approx 250 - 750$ keV) in $^{21}$Ne. However key properties of the most important states have not been measured. These unknown properties lead to large uncertainties in the reaction rates though no previous rate estimates have provided uncertainties \citep{PhysRevC.87.045805,CAUGHLAN1988283}. Here, a new calculation of the $^{17}$O$+\alpha$ reaction rates is presented, providing realistic uncertainties for the first time. The rates are derived from a rigorous evaluation of excited states in $^{21}$Ne based on a new measurement of the $^{20}$Ne($d,p$)$^{21}$Ne reaction, together with an evaluation of existing literature.

\section{Existing literature on $^{21}$Ne levels}

Some experimental data on levels in $^{21}$Ne are available from studies using a variety of populating reactions. As these works are used in the present paper to constrain the properties of observed levels, they are briefly summarised.

Direct measurements of $^{17}$O($\alpha,\gamma$)$^{21}$Ne \citep{PhysRevC.83.052802,TAGGART2019134894,WilliamsPRCSubmitted} and $^{17}$O($\alpha,n$)$^{20}$Ne \citep{PhysRevC.87.045805} have been performed but were not able to observe all states within the Gamow window due to the prohibitively low cross-sections.
The $^{17}$O($\alpha,\gamma$)$^{21}$Ne and $^{17}$O($\alpha,n$)$^{20}$Ne measurements in forward kinematics of \cite{PhysRevC.83.052802,PhysRevC.87.045805} used anodised tantalum-oxide targets with enriched $^{17}$O and a germanium detector. Measurements of the $^{17}$O($\alpha,\gamma$)$^{21}$Ne radiative-capture reaction were performed in inverse kinematics with the DRAGON recoil separator by \cite{TAGGART2019134894} and \cite{WilliamsPRCSubmitted}.

The $^{17}$O($\alpha,\gamma$)$^{21}$Ne measurement of 
\cite{PhysRevC.83.052802} observed three resonances at $E_r = 811$, $1122$, and $1311$ keV. The subsequent study of the same reaction by 
\cite{TAGGART2019134894} observed resonances at $E_r = 633$, $721$, $810$ and $1122$ keV. Some of the resonance strengths were revised by \cite{WilliamsPRCSubmitted} based on new DRAGON measurements with higher beam intensities, including an upper limit for the strength of the $E_r = 612$-keV resonance in $^{17}$O($\alpha,\gamma$)$^{21}$Ne.

\cite{PhysRevC.87.045805} also measured the $^{17}$O($\alpha,n_\mathrm{tot}$)$^{20}$Ne reaction using tantalum-oxide targets. The neutrons were detected in 20 $^3$He counters within a polyethylene moderator. The $^{17}$O($\alpha,n_1\gamma$)$^{20}$Ne reaction was measured by observation of $\gamma$ rays depopulating the first-excited state of $^{20}$Ne. The excitation function was measured between $E_{cm} = 650$ and $1860$ keV. A number of resonance structures were observed and an $R$-matrix analysis was performed. Rates of the $^{17}$O($\alpha,\gamma$)$^{21}$Ne and $^{17}$O($\alpha,n$)$^{20}$Ne reactions were reported. These rates included estimated contributions from unmeasured resonances below the region scanned in the excitation function, making assumptions about unknown spins and parities in combination with known information about levels in $^{21}$Ne \citep{ENSDF}. 

A subsequent uncertainty analysis of the measured $^{17}$O($\alpha$,n)$^{20}$Ne excitation functions presented in the literature was performed by \cite{Mohr2017}, which raised some doubts about the consistency of the data presented in \cite{PhysRevC.87.045805} in comparison to historical literature. However, the study found the largest discrepancy was in the high-energy cross section affecting the reaction rate at temperatures above 1 GK. At those energies, \cite{Mohr2017} suggested that the cross section presented in \cite{PhysRevC.87.045805} should be lowered by a factor of 2-3. At lower energies, though, their data appears to be in better agreement with the literature (see Fig. 5 of \cite{Mohr2017}). Furthermore, at the temperatures of interest in the present work (0.2-0.3 GK), the reaction rate is dominated by un-observed resonances that we treat in a rigorous manner and were not considered in \cite{Mohr2017}, which focused on applying the statistical model to compute reaction rates.

Additional key information about the excited states has been extracted from earlier measurements, including scattering and transfer studies.
\cite{PhysRev.114.194} performed a neutron resonance-scattering experiment. Neutrons were produced by bombarding a zirconium tritide and $^7$Li targets with protons. The incident neutron energies were varied by changing the bombarding energy of the proton beam. Information on excitation energies, widths, and spins and parities for $^{21}$Ne levels above the neutron threshold was determined. The uncertainty in the neutron energy scale was less than 5 keV and the neutron energy resolution was around 13 keV. Resonances with widths as small as $\Gamma = 6(2)$ keV were observed in this experiment.

Various $\gamma$-ray studies using reactions such as $^{18}$O($\alpha,n\gamma$)$^{21}$Ne \citep{ROLFS1972641,1989_Hoffman}, $^{12}$C($^{13}$C,$\alpha$)$^{21}$Ne \citep{ANDRITSOPOULOS1981281,HALLOCK1975141} and $^{16}$O($^{7}Li,np$)$^{21}$Ne \citep{Thummerer_2003,Wheldon_2005} provided additional information. Spins and parities of excited levels were assigned on the basis of observed decay branching and angular distributions. The observation of $\gamma$-ray decays from neutron-unbound levels in $^{21}$Ne can be used in combination with other observables to rule out certain spins and parities since the neutron partial width of a state cannot exceed the $\gamma$-ray partial width by a significant factor if a $\gamma$-ray transition is observed depopulating the state.

The  $^{20}$Ne($d,p$)$^{21}$Ne single-neutron transfer reaction has previously been studied by \cite{stanford1980study}. In this experiment, the differential cross sections and analysing powers of eight states in $^{21}$Ne were measured using a vector polarised deuteron beam on an enriched $^{20}$Ne gas cell target. The resolution was around 100 keV, largely due to the significant energy loss through the target for the deuteron beam. Large deviations between the calculated differential cross sections from the DWBA and the experimental data were observed at higher angles. This is likely due to the strong deformation of the $^{20}$Ne core. Inelastic excitation of deformed nuclei means that treatment with the DWBA may no longer accurately reflect the observed differential cross section (e.g. the $^{20}$Ne($d,^3$He)$^{19}$F study of \cite{DUDEK1971309}. For this reason, analysis of the strength of the $^{20}$Ne($d,p$)$^{21}$Ne single-neutron transfer reaction should be limited to centre-of-mass angles below around 30 degrees.

\section{Experimental details}

The experiment was performed at the Triangle Universities Nuclear Laboratory (TUNL). Deuterons 
were accelerated to 14~MeV through the tandem accelerator 
with an energy precision of better than 1~keV. Typical 
currents recorded with an electron-suppressed beamstop downstream from the target were 300-575~nA, except at the most forward angle where currents were limited to around 90~nA.  

The target consisted of $^{20}$Ne implanted into a 44~$\mu$g$/$cm$^2$ carbon foil with a Ne/C abundance ratio of 4.3$\pm$0.3$\%$ determined by  Rutherford Backscattering Spectrometry. 
The neon content was monitored with deuteron elastic scattering at $\theta_{lab}$ = 25\textdegree\ after collecting $^{20}$Ne($d,p$)$^{21}$Ne data at each angle to account for any target degradation. 
Uncertainties in neon target areal density were of the order of 7$\%$.\\ 
Reaction products entered the TUNL Enge split-pole spectrograph through a 0.54$\pm$0.01\ msr aperture and were momentum analyzed at the focal plane. The focal-plane detector comprised of two position-sensitive gas avalanche counters, a $\Delta$E proportional counter, and a scintillation counter~\citep{marshall2018focal}. Protons were identified using a $\Delta$E-position cut. Data were collected at five laboratory angles: 10\textdegree, 15\textdegree, 20\textdegree, 25\textdegree\ and 38\textdegree. Additional data for background characterisation were collected using a natural carbon target at each angle.

A quadratic internal calibration using the $^{21}$Ne states at $E_\mathrm{x} = 6609(1)$, $7420(1)$, $8069(2)$ and $8189(2)$~keV \citep{ENSDF} was used to convert focal-plane position to excitation energy. The Bayesian method outlined by \cite{marshall2018focal} gave realistic $E_\mathrm{x}$ uncertainties, explicitly including the statistical uncertainties in the fitted peak centroids and the systematic uncertainties from the focal-plane calibration. For peaks observed at multiple angles, our recommended energy was obtained from a weighted average of individual measurements. The reported energy uncertainty was conservatively constrained to be no smaller than the uncertainty at a single angle.

\begin{figure}
    \centering
    \includegraphics[width=0.48\textwidth]{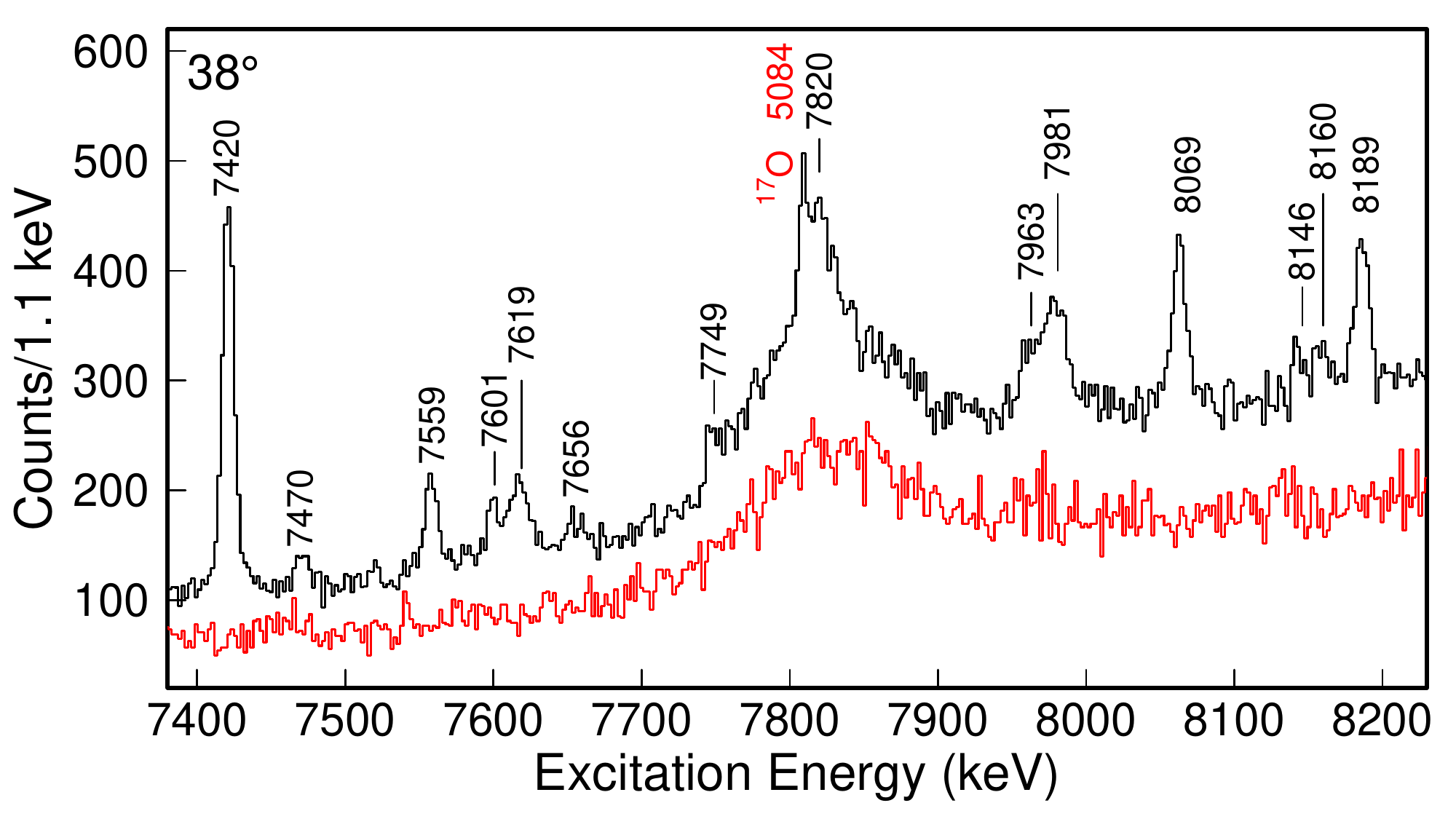}
    \caption{A proton focal-plane position spectrum from the $^{20}$Ne($d,p$)$^{21}$Ne reaction at $\theta_{lab}$ = 38\textdegree. Peaks are labelled with their corresponding excitation energy in $^{21}$Ne (keV). The carbon spectrum (arbitrary scale) is shown in red, highlighting the location of the $^{17}$O background contamination peak.}
    \label{fig:FPSpectrum}
\end{figure}

An example focal-plane spectrum for the astrophysically important region is shown in Fig.~\ref{fig:FPSpectrum}.
To extract peak intensities and positions, the focal plane was divided into regions based on the behaviour of the background reactions determined from the carbon target. For more details on the peak fitting see \cite{JFSThesis}. 
Angular distributions were extracted from the yields for each state observed at each angle accounting for beam on target, aperture solid angle, target content and dead time, which was typically below 9$\%$.

\section{Calculation of neutron widths}

The code {\sc fresco} \citep{thompson1988coupled} was used to calculate differential cross-sections under the assumption of a single-step reaction, in order to assign the transferred angular momentum, $\ell_n$, and extract the neutron width, $\Gamma_n$. Calculations were performed using the first-order distorted-wave Born approximation since at these low energies the breakup effects on the deuteron in the entrance channel are minimal \citep{PhysRevC.1.976}. Past $^{20}$Ne($d,p$)$^{21}$Ne transfer reactions reported in \cite{stanford1980study} have observed significant deviation between the expected differential cross section and the observed data at angles higher than $\theta_{lab} = 25$\textdegree\ degrees. For this reason, $\theta_{lab} = 38$\textdegree\ data was not used to assign $\ell_n$. The $\theta_{lab} = 38$\textdegree\ data were used to determine $E_\mathrm{x}$ and constrain the total width for observed states.

\begin{figure}
    \centering
    \includegraphics[width=0.48\textwidth]{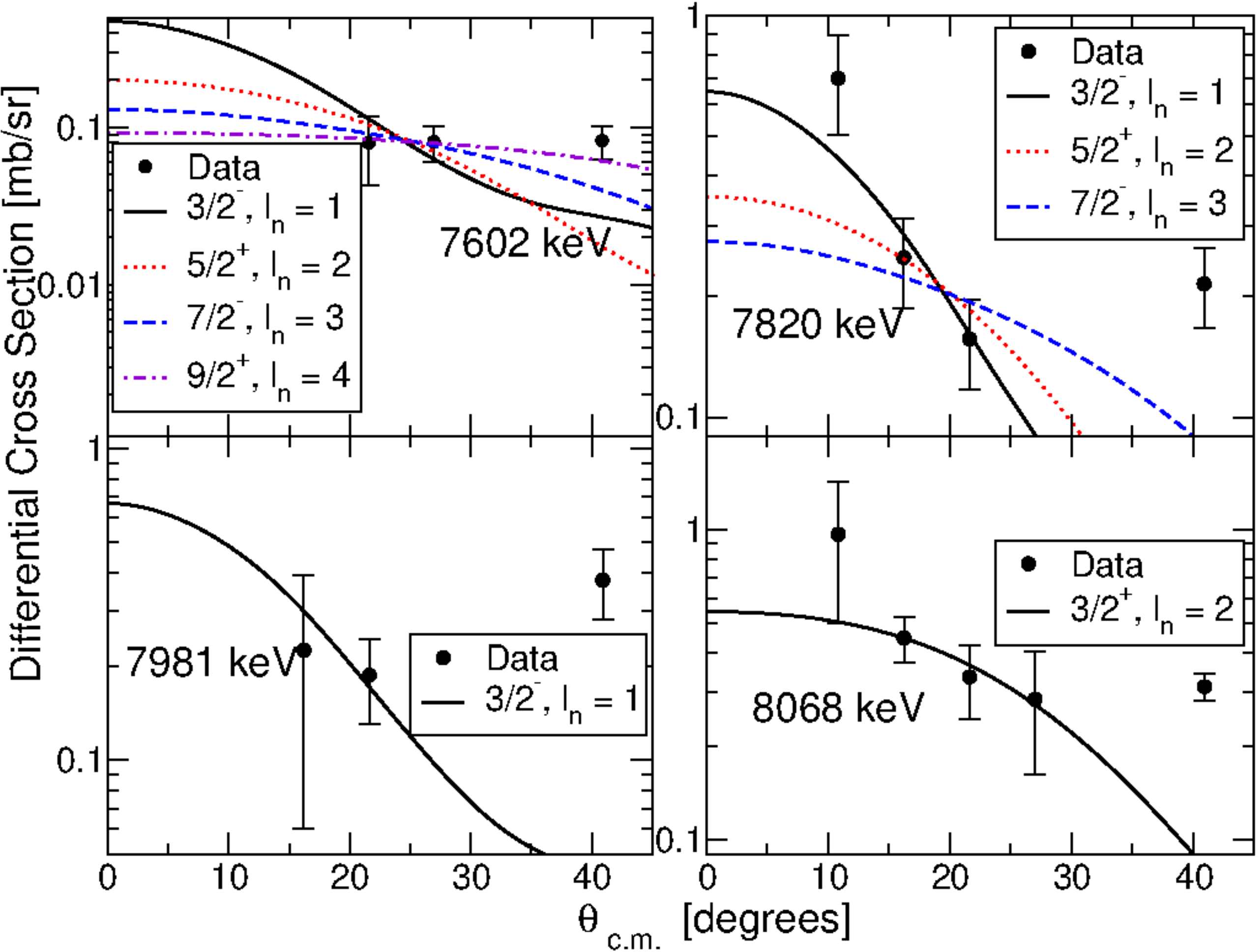}
    \caption{Experimental differential cross sections for some of the relevant states discussed in the present work. The $7981$- and $8068$-keV states have known $J^\pi$ \protect\citep{PhysRev.114.194,ENSDF} while the $7602$- and $7820$-keV states have unknown $J^\pi$. Note that, while the 38\textdegree\ data are included in the plot, they are not used in the DWBA analysis.}
    \label{fig:DWBAExample}
\end{figure}

The optical model potentials for $^{20}$Ne$+d$, $^{21}$Ne$+p$, $^{20}$Ne$+n$, $n+p$ and $^{20}$Ne$+p$ were taken from \cite{2006_An}, \cite{1991_Varner}, \cite{1997_Madland},  \cite{10.1143/PTPS.89.32} and \cite{1971_Menet}, respectively. Spectroscopic factors ($C^2S$) relating the DWBA and experimental cross sections:
\begin{equation}
    \frac{d\sigma}{d\Omega}_\mathrm{exp} = C^2S \frac{d\sigma}{d\Omega}_\mathrm{DWBA}
\end{equation}
were found by normalising the calculations to the experimental data. Example differential cross sections for four states are shown in Fig. \ref{fig:DWBAExample}. Two states with known $J^\pi$ (the $J^\pi = \nicefrac{3}{2}^-$, $E_\mathrm{x} = 7981$-keV state and $J^\pi = \nicefrac{3}{2}^+$, $8068$-keV state) have been included to demonstrate that the DWBA calculations reproduce the data well. Two states with inconclusive $J^\pi$ assignments are also shown.

The wave-functions from {\sc fresco}, $\phi(r)$, were used to compute the neutron widths, $\Gamma_n$ as in \cite{ILIADIS1997166}:
\begin{equation}
	\Gamma_n = 2 P_\ell(E_n,a) \frac{\hbar^2a}{2\mu}\,C^2S\,|\phi(a)|^2,
	\label{eqn:Spectroscopic factor}
\end{equation}
where $P_\ell(E_n,a)$ is the penetrability of a neutron of energy $E_n$ and orbital angular momentum $\ell$ evaluated at the radius $a$; $\mu$ is the reduced mass. The radius, $a$, was chosen to be where the $^{20}$Ne$+n$ wave-function is at 99\% of the asymptotic value \citep{PhysRevC.102.035803} and varied for different binding energies and $\ell_n$. We used the weak-binding approximation, calculating at various positive binding energies and extrapolating with a quadratic function to the negative, physical neutron binding energy \citep{PhysRevC.102.035803}. The uncertainties resulting from the calculations are much smaller than the uncertainty in the absolute normalisation from other sources.

\section{Evaluation of $^{17}$O$+\alpha$ reaction rates}

Table \ref{tab:Ne21Levels} summarises the spectroscopic information on relevant states in $^{21}$Ne above the $\alpha$-particle threshold. We discuss below the spin-parity assignments of astrophysically important states from which we produce physically-motivated reaction rates, with uncertainties for the first time. A comprehensive discussion on all of the states observed in this work will be presented in a forthcoming paper. 

The $E_\mathrm{x} = 7559$-, $7820$-, $8146$- and $8189$-keV states all have differential cross sections which are consistent with $\ell_n = 1$ or $2$ assignments. For $\ell_n = 1$ the $\Gamma_n$ are, for $J^\pi = \nicefrac{1}{2}^-$, $25(1)$, $20(3)$, $19(5)$ and $130(13)$, respectively (for $J^\pi = \nicefrac{3}{2}^-$, $\Gamma_n = 14(1)$, $11(2)$, $11(3)$, and $74(7)$ keV, respectively). Resonances with $\Gamma_n$ of this size would both have been observed in the neutron-scattering study reported in \cite{PhysRev.114.194} and would additionally result in visible broadening of the states in the focal-plane spectrum. Since neither of these are observed, $\ell_n = 1$ is ruled out for all of these states.

The differential cross section of the $E_\mathrm{x} = 7602$-keV state is consistent with $\ell_n = 2-4$. This state has been observed in $\gamma$-ray data of \cite{ROLFS1972641}. 
An $\ell_n=2$ assignment results in a neutron width of more than 100 eV. Based on lifetimes reported in that work, this neutron width greatly exceeds realistic $\gamma$-ray partial widths. It is, therefore, unlikely that $\gamma$-ray decay from this state would be observed if the $\ell_n=2$ assignment were made. We therefore assign this level $\ell_n = 3$ or $\ell_n=4$.

Close to $E_\mathrm{x} = 7982$ keV are two states, one narrow at $E_\mathrm{x} = 7982.1(6)$ keV \citep{ENSDF,TAGGART2019134894}, likely not observed in this experiment, and a broader one at $E_\mathrm{x} = 7981(2)$ keV, likely the state populated in the present work. 
The neutron width ($\Gamma_n = 6(2)$ keV), spin and parity ($J^\pi = \nicefrac{3}{2}^-$) of the $E_\mathrm{x} =7981(2)$-keV state are known from $^{20}$Ne$+n$ resonance scattering  \citep{PhysRev.114.194}. From the current data, we determine $\Gamma_n = 14(5)$ keV.

%
\begin{landscape}
\begin{table}
   \centering
    \caption{Spectroscopic information for relevant states above the $\alpha$-particle threshold in $^{21}$Ne. Previous excitation energies from other sources are given in the second column. The neutron partial widths ($\Gamma_n$) are those determined in the present experiment. Widths given in bold are measured or experimentally constrained values. The $\Gamma_\gamma = 0.20(14)$ eV are taken from the average of measured lifetimes in \protect\cite{ROLFS1972641} except where noted in the final column to preserve the $\Gamma_n/\Gamma_\gamma$ ratio of  \protect\cite{PhysRevC.87.045805} for resonances for which no updated information is available. Resonance information on higher resonances is taken from  \protect\cite{PhysRevC.87.045805}.}
    \def\arraystretch{2}
    \begin{tabular}{c | c | c | c | c | c | c | c | c}
    \hline 
        \makecell{$E_\mathrm{x}$\\\ [keV]} & \makecell{Previous $E_\mathrm{x}$\\\ [keV]} & \makecell{$E_{r,\alpha}$\\\ [keV]} & $2J^\pi$ & $\ell_\alpha$ & $\ell_n$ & \makecell{$\Gamma_\alpha$\\\ [eV]} & \makecell{$\Gamma_n$\\\ [eV]} & 
        Comments \\ \hline 
        
        $7420.4(15)^* $ & $7420.3(10)$ & $72.5(15)$ & $(5,7)^-$ & $1$ & $3$ & $1.2\times10^{-33}$ & $\mathbf{14(1),11(1)}$ & 
     \\
        \hline
        $7470(2)$    & $7465(10)$ & $122(2)$   & $(1,3)^-$ & $3,1$ & $1$ & $7.9\times10^{-24}$, $3.9\times10^{-22}$ & $200(140)$ & \makecell{Adopt $\frac{\Gamma_\gamma}{\Gamma_n} = 10^{-3}{}^a$} \\
        \hline
        $7559.1(15)$ & $7547(10)$ & $211.2(15)$ & $(3,5)^+$ & $2,0$ & $2,2$ & $2.4\times10^{-14}$, $2.5\times10^{-13}$ & $\mathbf{570(30)}$, $\mathbf{420(20)}$ &  \\
        \hline
        $7602.0(15)$ & $7600(5)$ & $254.1(15)$ & \makecell{$(5,7)^-$\\$(7,9)^+$} & \makecell{$1$\\$2$} & \makecell{$3$\\$4$} & \makecell{$2.6\times10^{-11}$\\$5.6\times10^{-12}$} & \makecell{$\mathbf{8(2)}$, $\mathbf{6(2)}$ \\ $\mathbf{0.4(1)}$, $\mathbf{0.3(1)}$} &  \\
        \hline
        $7619.9(10)$ & $7628(10)$ & $272.0(10)$ & $3^-{}^f$ & $1$ & $1$ & $1.7\times10^{-10}$ & $\mathbf{8000(1000)}$ &  \\
        \hline
        $7655.7(22)$ & $7648(2)$ & $307.8(22)$ & $7^+{}^h$  & $2$ & $4$ & $9.8\times10^{-10}$ & $0.10(7)$ & \makecell{Adopt $\frac{\Gamma_\gamma}{\Gamma_n} = 2^a$} \\
        \hline
        $7748.8(17)$ & $7740(10)$ & $400.9(17)$ & $5^+{}^a$ & $0$ & $2$ & $5.2\times10^{-6}$ & $200(140)$ & \makecell{Adopt $\frac{\Gamma_\gamma}{\Gamma_n} = 10^{-3}{}^a$} \\
        \hline
        $7820.1(15)$ & $7810(10)$ & $472.2(15)$ & $(3,5)^+$ & $2,0$ & $2$ & $1.8\times10^{-5}$, $1.7\times10^{-4}$ & $\mathbf{560(90)}$, $\mathbf{400(60)}$ &  \\
        \hline
        $7960(2)$    &  $7960.9(13)$ & $612(2)$ & $11^-{}^g$ & $3$ & $5$ & $5.3\times10^{-8}$ & $0.10(7)$ & \makecell{Adopt $\frac{\Gamma_\gamma}{\Gamma_n} = 2^a$} \\
        \hline
        $7981(2)$    & $7980(10)$ & $633(2)$ & $3^-{}^f$  & $1$ & $1$ & $1.9\times10^{-2}$ & $\mathbf{14000(5000)}$ &  \\
        \hline
        $7982.1(7)$ & $7982.1(6)$ & $634.2(7)$    &                    &     &     & {\bf $7.5(15) \times 10^{-6}$} & & \makecell{From $\omega\gamma_{(\alpha,\gamma)} = 4.98(97)$ $\mu$eV$^e$} \\
        \hline
        $8008(2)$    & $8009(10)$ & $660(2)$    & $1^-{}^f$ & $3$ & $1$ & $1.2\times10^{-3}$ & $0.20(14)$ &  $\Gamma_\gamma = 2.0(14)\times10^{-4}$ eV \\
        \hline
             $8069(1)^* $ & $8069(2)$  & $721(1)$ & $3^+{}^a$ & $2$ & $2$ & $\mathbf{46.2(46)\times10^{-3}}$ & $\mathbf{1600(200)}$ & $\Gamma_\gamma = 0.54(35)$ eV$^d$\\
        \hline      
             $8146(1)$ & $8146(2)$ & $798(1)$ & $3^+{}^a$ & $2,0$ & $2$ & $\mathbf{54.7(55)\times10^{-3}}$ & $\mathbf{550(150)}$, $\mathbf{400(100)}$ &  \\
        \hline      
              $8159(2)^b$ & $8155.0(10)$ & $811(2)$ & $9^+{}^g$ & $2$ & $4$ & & & $\omega\gamma_{(\alpha,,\gamma)} = 7.72(55)^c$ meV  \\
        \hline
              $8160(2)$ & $8160(2)$ & $812(2)$ & $5^+{}^a$ & $0$ & $2$ & $\mathbf{1.6(2)\times10^{-3}}$ & $\mathbf{23000(2300)}$ & \makecell{Adopted from\\Ref. \cite{WilliamsPRCSubmitted}}\\ 
              \hline
    \end{tabular}
   \label{tab:Ne21Levels}
   \raggedright \\
   {$^*:$ used in calibration}\\
   {$a:$ \cite{PhysRevC.87.045805}}\\
   {$b:$ \cite{PhysRevC.83.052802}}\\
   {$c:$ Weighted average of \cite{WilliamsPRCSubmitted} and \cite{PhysRevC.83.052802}.}\\
   {$d:$ To preserve the measured $\omega\gamma_{(\alpha,\gamma)}$ from Ref. \cite{WilliamsPRCSubmitted}.}\\
   {$e:$ \cite{WilliamsPRCSubmitted}}\\
   {$f:$ \cite{PhysRev.114.194}}\\
   {$g:$ \cite{Thummerer_2003}}\\
   {$h:$ \cite{1989_Hoffman}}
\end{table}
\end{landscape}

Reaction rates were calculated using the {\sc RatesMC} Monte Carlo code \cite{2010NuPhA.841....1L,Sallaska_2013}. Data on directly measured $^{17}$O($\alpha,\gamma$)$^{21}$Ne and $^{17}$O($\alpha,n$)$^{20}$Ne resonances were taken from  \cite{PhysRevC.83.052802,PhysRevC.87.045805,WilliamsPRCSubmitted}. The weighted mean with inflated uncertainties described by \cite{PhysRevC.85.065809} was used for the $^{17}$O($\alpha,\gamma$)$^{21}$Ne resonance strengths to account for systematic uncertainties. For many of the resonances listed in \cite{PhysRevC.87.045805} there are no listed uncertainties. These were arbitrarily assumed to be 10\% for all widths. For any resonances for which all partial widths could be estimated, the reaction rate was numerically integrated. 

For resonances with no measured $\Gamma_\alpha$, the reduced $\alpha$-particle width was sampled from a Porter-Thomas distribution \citep[see][for details]{Sallaska_2013,PhysRevC.85.065809}. The neutron widths were, where available, taken from the present work. $\Gamma_\gamma = 0.20(14)$ eV was used, from the average of the lifetimes in \cite{ROLFS1972641}. For resonances where the $\Gamma_n$ and $\Gamma_\gamma$ are unknown, the ratio of these widths is adopted from \cite{PhysRevC.87.045805}. 
Importantly, we adopt the tentative $J^\pi = \nicefrac{7}{2}^+$ assignment for the $E_r = 308$-keV resonance from \cite{ENSDF} which reduces the contribution of this state to the $^{17}$O($\alpha,\gamma$)$^{21}$Ne reaction rate compared to \cite{PhysRevC.87.045805} in which a $J^\pi = \nicefrac{5}{2}^+$ assignment was used. We furthermore note that a $\nicefrac{5}{2}^+$ assignment is inconsistent with systematic trends in $^{21}$Ne in which no other $J^\pi = \nicefrac{5}{2}^+$ states at a similar excitation energy decay by $\gamma$-ray emission.

The $E_\mathrm{x} = 7749$-keV state was only observed at one angle due to $^{16}$O contamination. The peak observed in the present experiment does not appear in the background spectrum and was therefore assumed to be a $^{21}$Ne state, but confirmation of this state with additional measurements is required. An angular distribution could not be extracted and the spin remains unconstrained. For the calculation of the rate we adopted the assumptions of \cite{PhysRevC.87.045805}.

\begin{figure}
    \centering
    \includegraphics[width=\columnwidth]{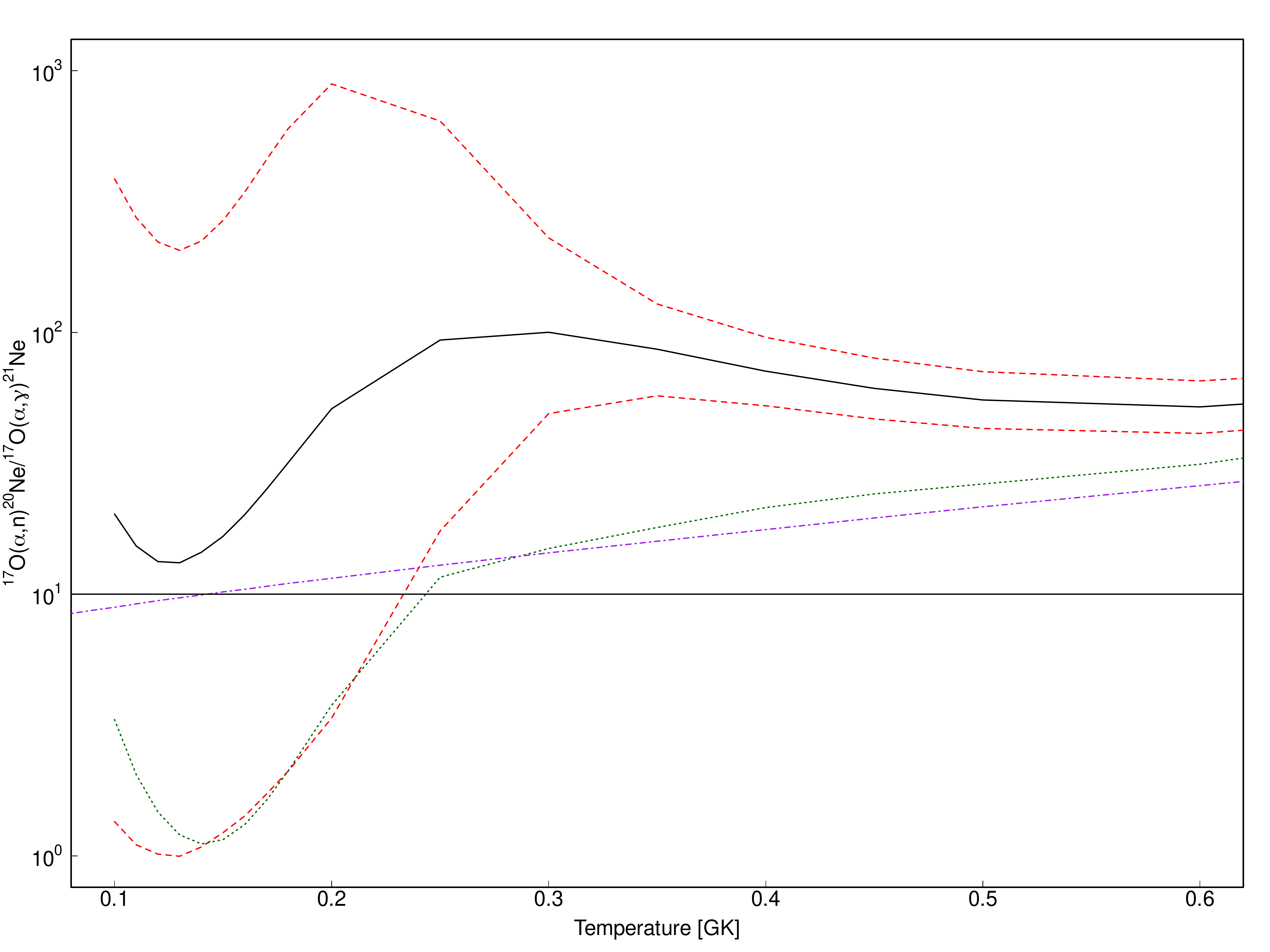}
    \caption{The ratios of the $^{17}$O($\alpha,n$)$^{20}$Ne and $^{17}$O($\alpha,\gamma$)$^{21}$Ne reaction rates. The solid black line shows the ratio of the median rates in the current calculation and the dashed red lines show the ratios of the upper (lower) $^{17}$O($\alpha,n$)$^{20}$Ne rate to the lower (upper) $^{17}$O($\alpha,\gamma$)$^{21}$Ne reaction rates. The green dotted line shows the ratio of the rates from \protect\cite{PhysRevC.87.045805} The purple dashed-dotted line shows the ratio of the rates from \protect\cite{CAUGHLAN1988283}.}
    \label{fig:RatioPlot}
\end{figure}

For the $^{17}$O($\alpha,\gamma$)$^{21}$Ne reaction, the Monte-Carlo calculations show that the dominant contributions are from the resonances at $E_\mathrm{r} = 308$, $634$ and $811$keV ($E_\mathrm{x} = 7656$, $7961$, $7982$ and $8159$ keV). The $634$- and $811$-keV resonances have measured strengths \citep{WilliamsPRCSubmitted, TAGGART2019134894,PhysRevC.83.052802}. For the $^{17}$O($\alpha,n$)$^{20}$Ne reaction, the dominant contributions are from the $E_\mathrm{r} = 401$-, $472$- and $721$-keV resonances ($E_\mathrm{x} = 7749$, $7820$ and $8069$ keV) with a small contribution from the $E_\mathrm{r} = 633$-keV resonance. The dominant contributions to the uncertainty in the reaction-rate ratio are the unknown $\Gamma_\alpha$ partial widths.

The $^{17}$O($\alpha,n$)$^{20}$Ne to $^{17}$O($\alpha,\gamma$)$^{21}$Ne reaction-rate ratio for our median rates is presented in Fig. \ref{fig:RatioPlot}. Also shown are $1\sigma$ \textquoteleft high\textquoteright\ and \textquoteleft low\textquoteright\ ratios from our work, and those from  \cite{CAUGHLAN1988283} and  \cite{PhysRevC.87.045805}.
The present ratio is significantly higher than that of Best {\textit{et al.}} between 0.25 and 0.7 GK for a number of reasons. There are some inconsistencies in the data in Table II of \cite{PhysRevC.87.045805} (e.g. the $E_r = 308$-keV resonance state), for which the listed resonance strengths are in disagreement with the $\Gamma_n$ and $\Gamma_\gamma$ ratio \citep{ABPrivateCommunication}. We have, additionally, changed spin-parity assignments where appropriate resulting in some changes in contributions of states to the reaction rates. Lastly, we have utilised the direct $^{17}$O($\alpha,\gamma$)$^{21}$Ne measurements from DRAGON \citep{WilliamsPRCSubmitted}.

\begin{figure}
    \centering
    \includegraphics[width=\columnwidth]{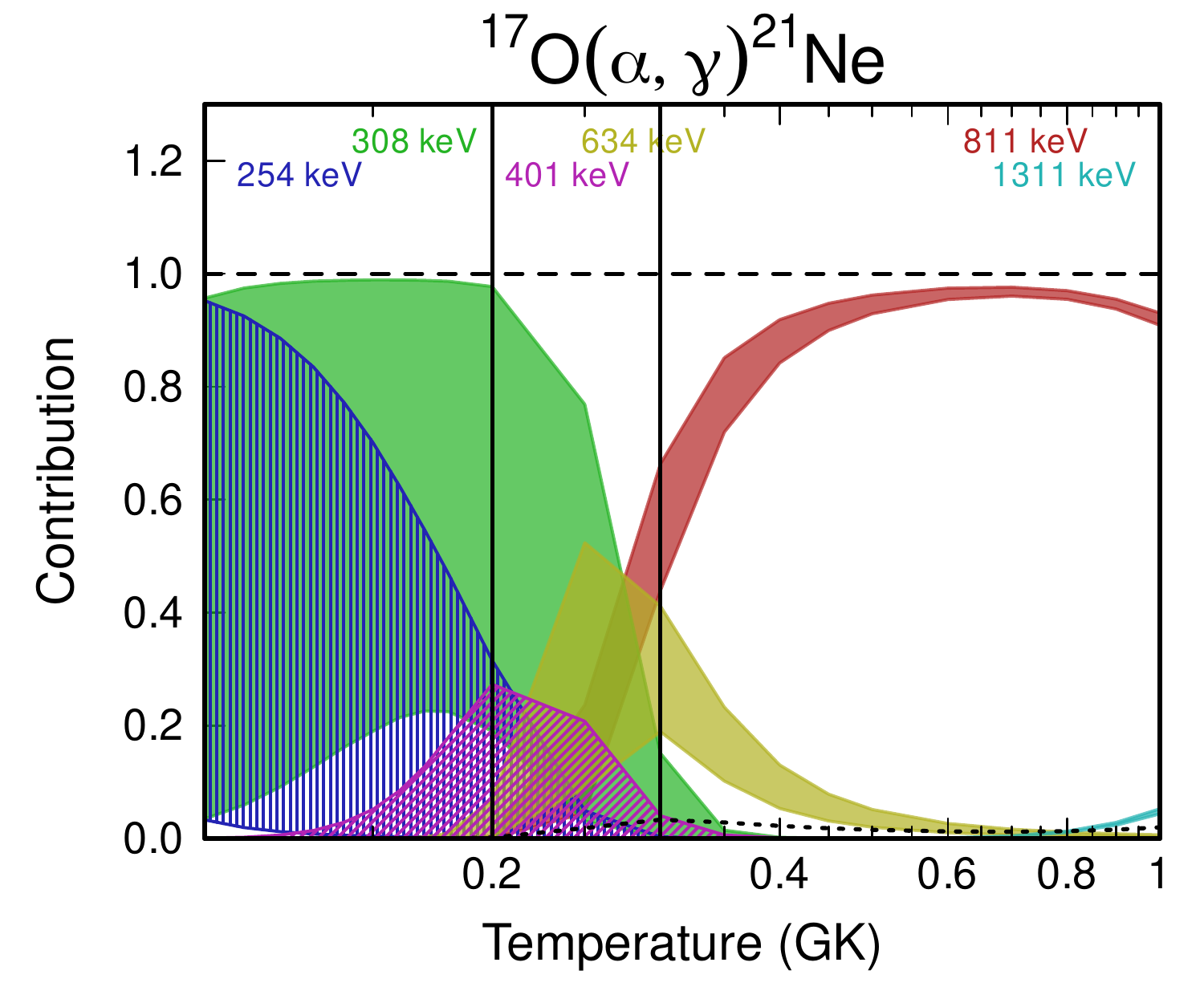}
    \includegraphics[width=\columnwidth]{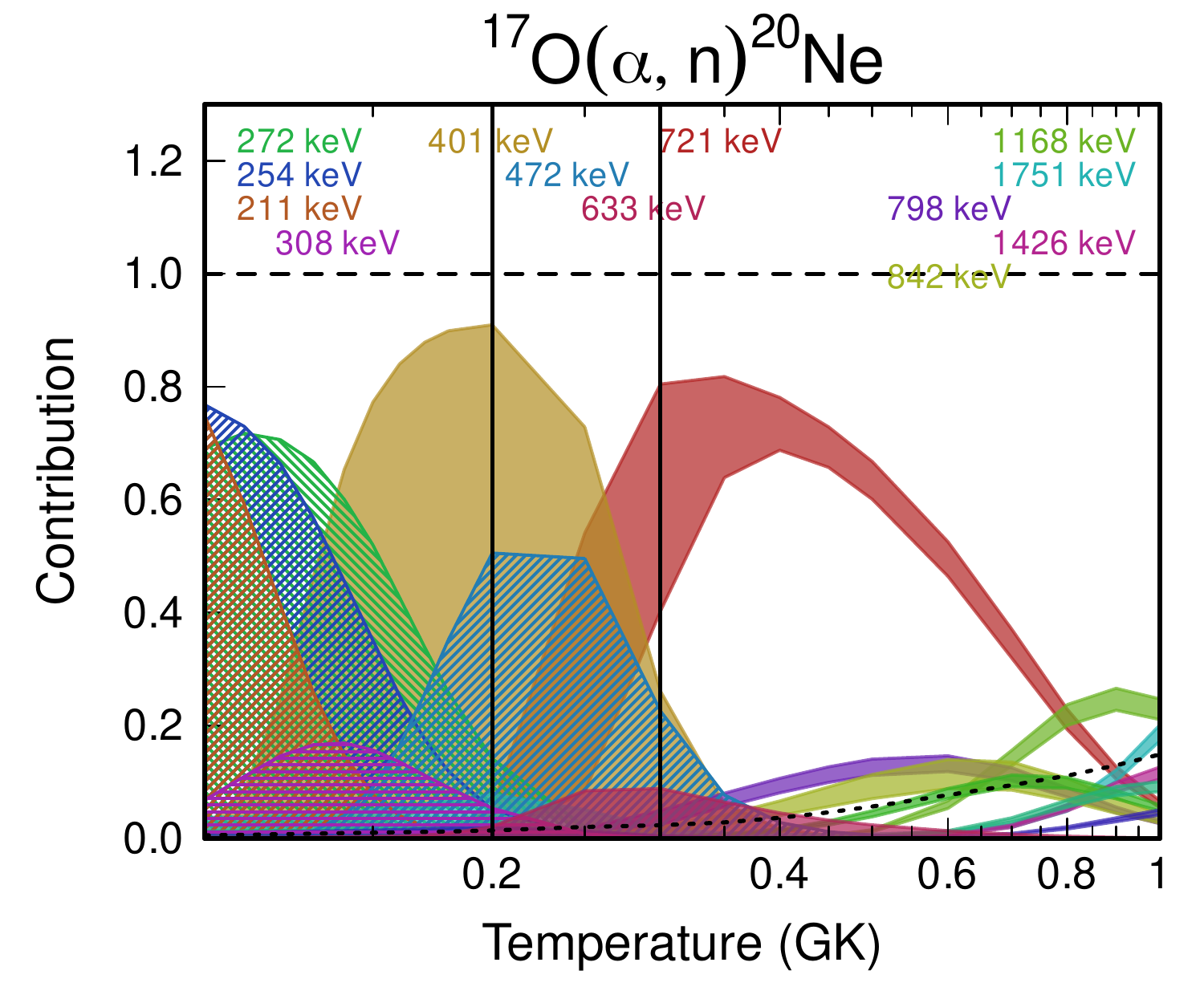}
    \caption{Contribution plots for (top) the $^{17}$O($\alpha,\gamma$)$^{21}$Ne reaction and (bottom) the $^{17}$O($\alpha,n$)$^{20}$Ne reaction indicating which resonances dominate the reaction rate within the Gamow window (vertical lines). The coloured bands show the uncertainty in the contribution of each resonance to the total rate.}
    \label{fig:contributions}
\end{figure}

Contribution plots for the two reactions are shown in Fig. \ref{fig:contributions} (see caption for details). For the $^{17}$O($\alpha,\gamma$)$^{21}$Ne reaction, the main contribution within the astrophysically relevant region is the $E_r = 308$-keV resonance, for which no estimate of the $\alpha$-particle width is yet available. More resonances can potentially contribute to the $^{17}$O($\alpha,n$)$^{20}$Ne reaction, since for many of these resonances the neutron partial width is known to be much larger than the $\gamma$-ray partial width. For most resonances, therefore, the $\gamma$-ray decay is vanishingly small and these states cannot meaningfully contribute to the flux of abundances through the $^{17}$O($\alpha,\gamma$)$^{21}$Ne reaction.

The $^{17}$O($\alpha,\gamma$)$^{21}$Ne and $^{17}$O($\alpha,n$)$^{20}$Ne reaction rates are shown in Tables \ref{tab:RR_ag} and \ref{tab:RR_an}, respectively.

\begin{table}
    \centering
    \caption{The $^{17}$O($\alpha,\gamma$)$^{21}$Ne reaction rate. The lower, median and upper columns correspond to the 32\%, 50\% and 68\% intervals for the computed reaction rate. The reaction rates are in units of cm$^3$ mol$^{-1}$ s$^{-1}$.}
    \def\arraystretch{1.5}
    \begin{tabular}{c c c c}
    $T$ [GK] & Lower & Recommended & Upper \\ \hline
        0.100 & 2.18$\times$10$^{-22}$ & 1.20$\times$10$^{-21}$ &
      5.33$\times$10$^{-21}$ \\ 
0.110 & 3.67$\times$10$^{-21}$ & 2.05$\times$10$^{-20}$ &
      9.19$\times$10$^{-20}$  \\ 
0.120 & 3.91$\times$10$^{-20}$ & 2.21$\times$10$^{-19}$ &
      1.01$\times$10$^{-18}$ \\ 
0.130 & 2.94$\times$10$^{-19}$ & 1.68$\times$10$^{-18}$ &
      7.95$\times$10$^{-18}$ \\ 
0.140 & 1.68$\times$10$^{-18}$ & 9.59$\times$10$^{-18}$ &
      4.66$\times$10$^{-17}$  \\ 
0.150 & 7.71$\times$10$^{-18}$ & 4.38$\times$10$^{-17}$ &
      2.18$\times$10$^{-16}$  \\ 
0.160 & 3.00$\times$10$^{-17}$ & 1.67$\times$10$^{-16}$ &
      8.43$\times$10$^{-16}$  \\ 
0.170 & 1.01$\times$10$^{-16}$ & 5.46$\times$10$^{-16}$ &
      2.81$\times$10$^{-15}$  \\ 
0.180 & 2.99$\times$10$^{-16}$ & 1.59$\times$10$^{-15}$ &
      8.22$\times$10$^{-15}$  \\ 
0.200 & 2.13$\times$10$^{-15}$ & 1.02$\times$10$^{-14}$ &
      5.13$\times$10$^{-14}$ \\ 
0.250 & 3.23$\times$10$^{-13}$ & 5.84$\times$10$^{-13}$ &
      1.71$\times$10$^{-12}$  \\ 
0.300 & 4.54$\times$10$^{-11}$ & 5.53$\times$10$^{-11}$ &
      6.94$\times$10$^{-11}$  \\ 
0.350 & 2.43$\times$10$^{-09}$ & 2.70$\times$10$^{-09}$ &
      3.02$\times$10$^{-09}$  \\ 
0.400 & 5.24$\times$10$^{-08}$ & 5.68$\times$10$^{-08}$ &
      6.18$\times$10$^{-08}$  \\ 
0.450 & 5.76$\times$10$^{-07}$ & 6.21$\times$10$^{-07}$ &
      6.69$\times$10$^{-07}$  \\ 
0.500 & 3.90$\times$10$^{-06}$ & 4.19$\times$10$^{-06}$ &
      4.50$\times$10$^{-06}$ \\ 
0.600 & 6.74$\times$10$^{-05}$ & 7.21$\times$10$^{-05}$ &
      7.71$\times$10$^{-05}$  \\ 
0.700 & 5.03$\times$10$^{-04}$ & 5.37$\times$10$^{-04}$ &
      5.72$\times$10$^{-04}$ \\ 
0.800 & 2.23$\times$10$^{-03}$ & 2.37$\times$10$^{-03}$ &
      2.53$\times$10$^{-03}$  \\ 
0.900 & 7.04$\times$10$^{-03}$ & 7.48$\times$10$^{-03}$ &
      7.95$\times$10$^{-03}$  \\ 
1.000 & 1.76$\times$10$^{-02}$ & 1.87$\times$10$^{-02}$ &
      1.98$\times$10$^{-02}$  \\ \hline
    \end{tabular}
    
    \label{tab:RR_ag}
\end{table}

\begin{table}
    \centering
    \caption{As Table \ref{tab:RR_ag} but for the $^{17}$O($\alpha,n$)$^{20}$Ne reaction.}
    \def\arraystretch{1.5}
    \begin{tabular}{c c c c}
    $T$ [GK] & Lower & Recommended & Upper \\ \hline
 0.100 & 7.04$\times$10$^{-21}$ & 2.52$\times$10$^{-20}$ &
      8.75$\times$10$^{-20}$  \\ 
0.110 & 9.98$\times$10$^{-20}$ & 3.24$\times$10$^{-19}$ &
      1.05$\times$10$^{-18}$ \\ 
0.120 & 1.01$\times$10$^{-18}$ & 3.00$\times$10$^{-18}$ &
      9.13$\times$10$^{-18}$  \\ 
0.130 & 7.92$\times$10$^{-18}$ & 2.22$\times$10$^{-17}$ &
      6.40$\times$10$^{-17}$  \\ 
0.140 & 5.02$\times$10$^{-17}$ & 1.38$\times$10$^{-16}$ &
      3.91$\times$10$^{-16}$  \\ 
0.150 & 2.66$\times$10$^{-16}$ & 7.30$\times$10$^{-16}$ &
      2.14$\times$10$^{-15}$  \\ 
0.160 & 1.20$\times$10$^{-15}$ & 3.40$\times$10$^{-15}$ &
      1.07$\times$10$^{-14}$ \\ 
0.170 & 4.75$\times$10$^{-15}$ & 1.42$\times$10$^{-14}$ &
      4.78$\times$10$^{-14}$ \\ 
0.180 & 1.70$\times$10$^{-14}$ & 5.30$\times$10$^{-14}$ &
      1.91$\times$10$^{-13}$ \\ 
0.200 & 1.70$\times$10$^{-13}$ & 5.58$\times$10$^{-13}$ &
      2.23$\times$10$^{-12}$ \\ 
0.250 & 3.63$\times$10$^{-11}$ & 7.30$\times$10$^{-11}$ &
      2.35$\times$10$^{-10}$ \\ 
0.300 & 4.72$\times$10$^{-09}$ & 5.78$\times$10$^{-09}$ &
      9.44$\times$10$^{-09}$  \\ 
0.350 & 1.92$\times$10$^{-07}$ & 2.15$\times$10$^{-07}$ &
      2.57$\times$10$^{-07}$ \\ 
0.400 & 3.24$\times$10$^{-06}$ & 3.54$\times$10$^{-06}$ &
      3.95$\times$10$^{-06}$ \\ 
0.450 & 2.99$\times$10$^{-05}$ & 3.23$\times$10$^{-05}$ &
      3.54$\times$10$^{-05}$ \\ 
0.500 & 1.80$\times$10$^{-04}$ & 1.94$\times$10$^{-04}$ &
      2.10$\times$10$^{-04}$ \\ 
0.600 & 2.90$\times$10$^{-03}$ & 3.08$\times$10$^{-03}$ &
      3.28$\times$10$^{-03}$  \\ 
0.700 & 2.46$\times$10$^{-02}$ & 2.59$\times$10$^{-02}$ &
      2.72$\times$10$^{-02}$  \\ 
0.800 & 1.49$\times$10$^{-01}$ & 1.55$\times$10$^{-01}$ &
      1.61$\times$10$^{-01}$ \\ 
0.900 & 7.29$\times$10$^{-01}$ & 7.56$\times$10$^{-01}$ &
      7.84$\times$10$^{-01}$ \\ 
1.000 & 3.03$\times$10$^{+00}$ & 3.13$\times$10$^{+00}$ &
      3.25$\times$10$^{+00}$ \\ \hline
    \end{tabular}
    \label{tab:RR_an}
\end{table}

\section{Astrophysical Implications}

\begin{figure}
    \centering
    \includegraphics[width=\columnwidth]{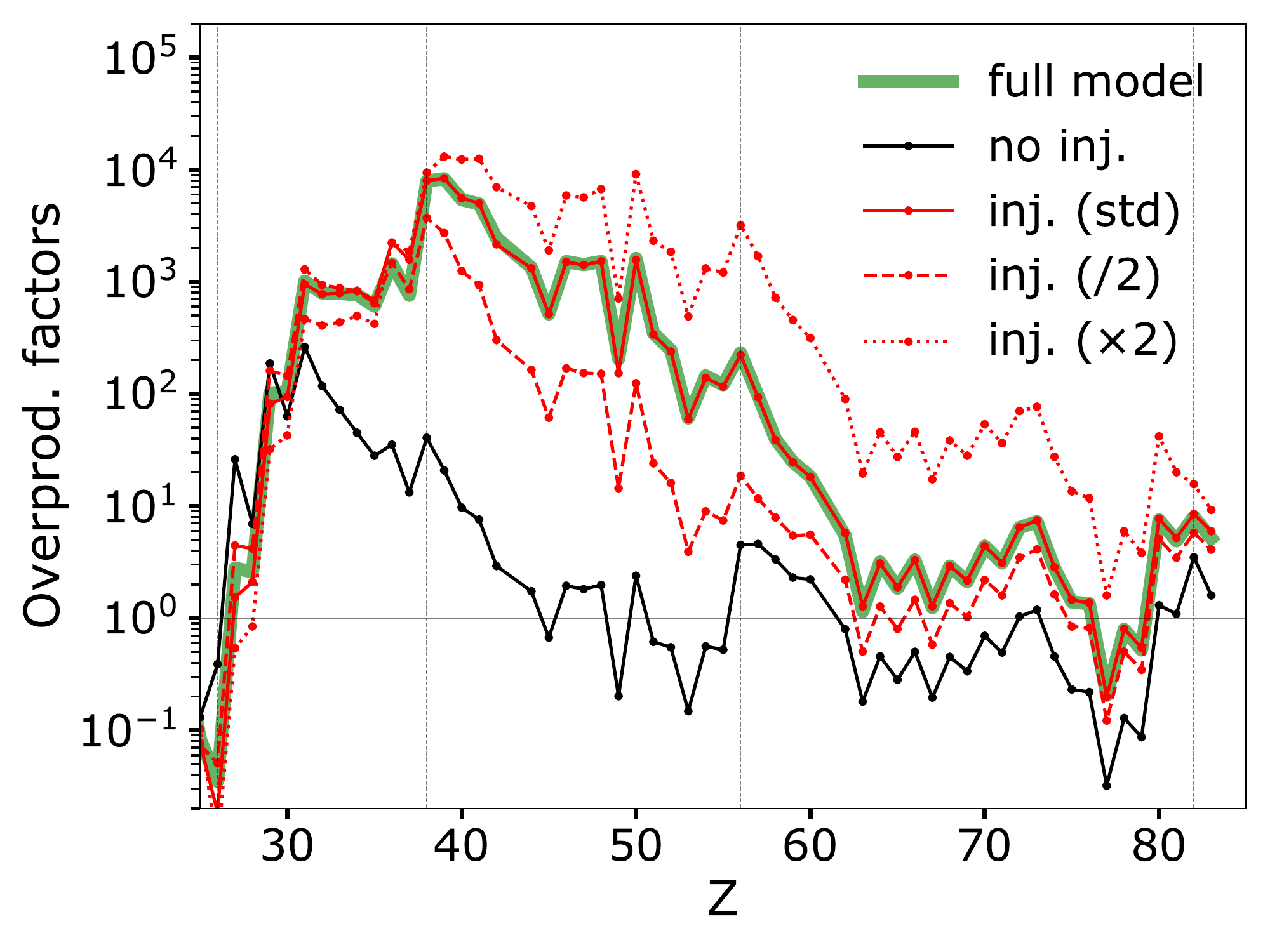}
    \caption{Overproduction factors of a one zone nucleosynthesis model mimicking the core helium burning phase of a rotating massive star at low metallicity. Different injection cases (see text for details) are considered: no injection (black pattern) and injection at three different rates (red patterns). The green pattern shows the outputs of the central layer of a full stellar model at the end of the core helium-burning stage. Here all models were computed with the rates of \protect\cite{PhysRevC.87.045805} for $^{17}$O($\alpha,\gamma$) and $^{17}$O($\alpha,n$). The 4 vertical lines highlight the elements Fe, Sr, Ba and Pb.}
\label{fig:Abundances}
\end{figure}

We tested the impact of the new $^{17}$O($\alpha,n$)$^{20}$Ne and $^{17}$O($\alpha,\gamma$)$^{21}$Ne rates on s-process nucleosynthesis using a simplified one-zone nucleosynthesis code mimicking core helium burning. 
Details on this code can be found in \cite{Choplin16}. The code was also used in \cite{Placco20} to make comparisons with an observed star enriched in trans-iron elements.
We follow the central temperature and density profiles obtained from a complete rotating 25~$M_{\odot}$ stellar model at a metallicity of $10^{-3}$ in mass fraction, computed with the Geneva stellar evolution code \citep{Eggenberger08}.
The initial composition of the one-zone code is extracted from the core of this stellar model, at core helium-burning ignition. 
To mimic rotation, $^{13}$C and $^{14}$N are injected (cf. Sect. \ref{sect:intro}) at a constant rate (expressed in $M_{\odot}$~yr$^{-1}$) during the nucleosynthesis calculation \citep{Choplin16}. 
During injection, $100$~times more $^{14}$N as $^{13}$C is injected as a typical value in full stellar models \citep[e.g. Fig.~9 in][]{Choplin18}. 
This factor of $\sim 100$ corresponds to the CNO $^{14}$N/$^{13}$C equilibrium ratio at $T \sim 80$~MK, which is found at the bottom of the H-burning shell in massive stars. 
The injection rate was calibrated so as to reproduce the central abundances of a full rotating stellar model at the end of the core helium-burning phase. 
For this calibration, we used the same rates as in the full stellar model, namely the rates of \cite{PhysRevC.87.045805} for $^{17}$O($\alpha,\gamma$)$^{21}$Ne and $^{17}$O($\alpha,n$)$^{20}$Ne, the rates of \cite{PhysRevC.85.065809} for $^{22}$Ne($\alpha,\gamma$)$^{26}$Mg and $^{22}$Ne($\alpha,n$)$^{25}$Mg and the rate from \cite{Guo12} for $^{13}$C($\alpha,n$)$^{16}$O. 
The adopted standard injection rate is $2.5\times 10^{-7} M_{\odot}$~yr$^{-1}$ and produces the overproduction factors\footnote{The overproduction factors are expressed as $X_{\rm f}/X_{\rm i}$ where $X_{\rm {i/f}}$ represent the initial/final mass fractions.} shown by the solid red line in Fig~\ref{fig:Abundances}. 
For most elements, it deviates by less than 5~\% from the full model (green pattern). 
Cases with higher (red dotted line) and lower (red dashed line) injection rates are shown for comparison. 
Without injection, we obtain a typical weak s-process pattern from non-rotating massive stars (black pattern).

\begin{figure*}
    \centering
    \includegraphics[width=\columnwidth]{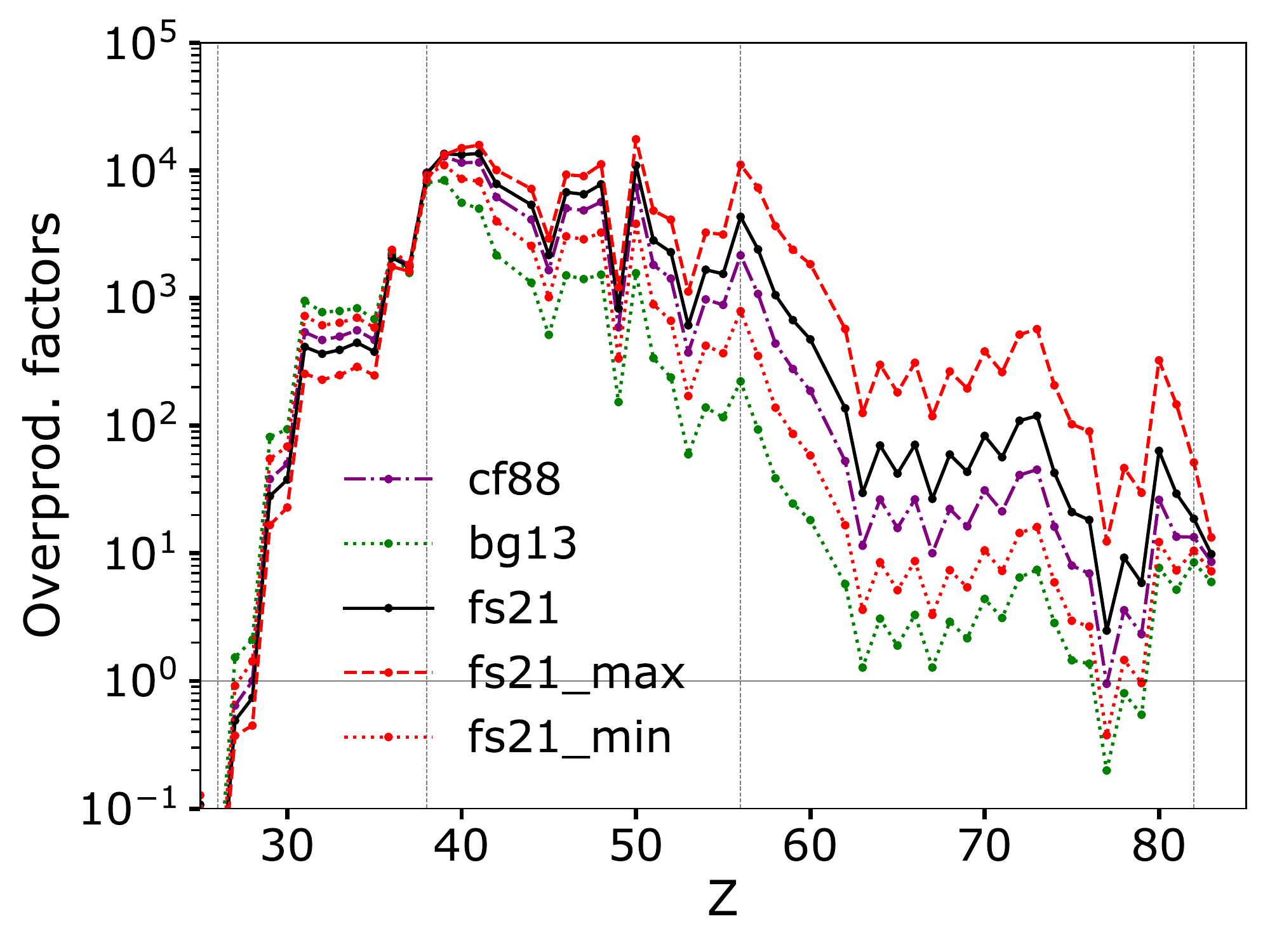}
    \includegraphics[width=\columnwidth]{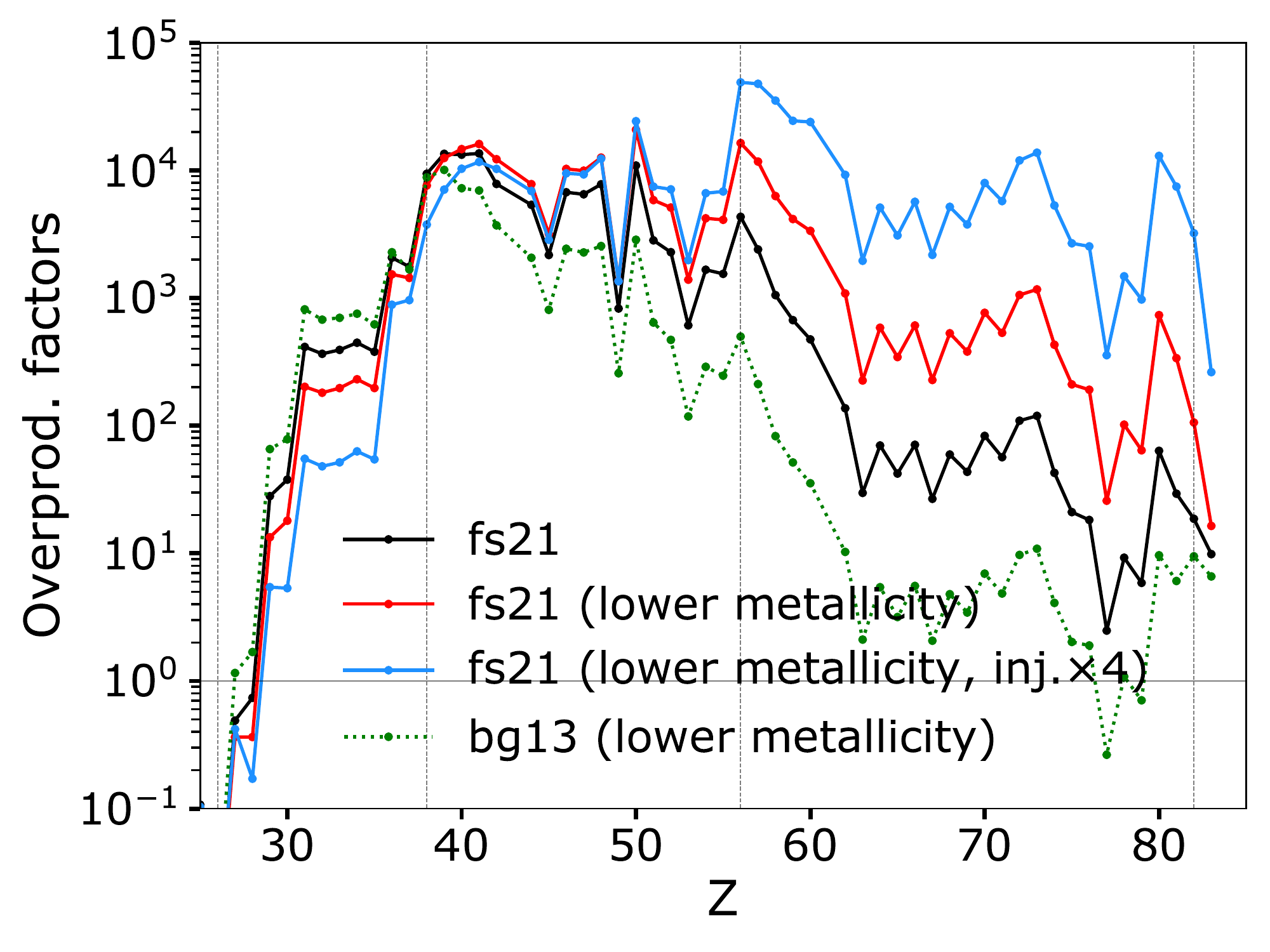}
    \caption{
    Overproduction factors of a one zone nucleosynthesis model mimicking the core helium burning phase of a rotating massive star \textbf{at low metallicity.}
    \textit{Left panel:} the five sets of rates shown in Fig.~\protect\ref{fig:RatioPlot} are tested: cf88  \citep{CAUGHLAN1988283}, bg13 \citep{PhysRevC.87.045805}, the recommended rates derived in this work (fs21) plus the two limiting cases shown in Fig.~\protect\ref{fig:RatioPlot}.
    The 4 vertical lines highlight the elements Fe, Sr, Ba and Pb.
    \textit{Right panel:} the fs21 model at a metallicity of $10^{-3}$ in mass fraction is shown again (black pattern) together with the same model but with initial abundances corresponding to a metallicity of $10^{-5}$ (red pattern). The blue model is computed like the red one but with an injection rate 4 times higher. The bg13 model at a metallicity of $10^{-5}$ is also shown.}
    \label{fig:Abundances2}
\end{figure*}

In the left panel of Fig.~\ref{fig:Abundances2}, all one-zone models were computed with the standard injection rate of $2.5\times 10^{-7} M_{\odot}$~yr$^{-1}$ obtained from the calibration discussed previously. The only differences between the models shown in Fig.~\ref{fig:Abundances2} (left panel) and the model shown by the solid red line in Fig.~\ref{fig:Abundances} are the rates of $^{17}$O($\alpha,n$)$^{20}$Ne and $^{17}$O($\alpha,\gamma$)$^{21}$Ne.
The recommended reaction rates were used (fs21), as well as the limiting cases of the minimum (fs21\_min) and maximum (fs21\_max) ($\alpha,n$)/($\alpha,\gamma$) ratio.
These results show that the s-process in rotating massive stars is likely to continue at least to barium, and potentially up to lead for the largest ($\alpha,n$)/$(\alpha,\gamma$) reaction-rate ratio. 
The scatter for elements with atomic number $Z>40$ goes up to about 2~dex. The bg13 and fs21\_min sets have the lowest $^{17}$O($\alpha,n$)$^{20}$Ne/$^{17}$O($\alpha,\gamma$)$^{21}$Ne ratios (Fig.~\ref{fig:RatioPlot}) hence giving the lowest yields (green and red dotted pattern). The cf88 and fs21 sets with higher ($\alpha,n$)/($\alpha,\gamma$) ratios substantially produce elements with $Z>55$ and the fs21\_max set shows the highest yields, as expected from the high ($\alpha,n$)/($\alpha,\gamma$) ratio. 

As an estimate of the impact of the new rate at very low metallicity, we include in the right panel of Fig.~\ref{fig:Abundances2} a one-zone calculation similar to the fs21 model (black pattern) but computed with an initial composition corresponding to a metallicity of $10^{-5}$ in mass fraction (red pattern). 
This shows that a lower initial metallicity combined with a similar neutron production leads to a stronger overproduction of elements heavier than atomic number $Z \sim 50$ at the expense of lighter elements.
This is even more visible if considering a higher injection rate (green pattern) which would correspond to a more efficient rotational mixing in full stellar models. 
A higher injection rate at lower metallicity is not unrealistic since rotational mixing is expected to be more efficient with decreasing metallicity \citep[e.g.][]{Maeder01}. 
We note that contrary to the fs21 rates, the bg13 rates at lower metallicity do not lead to significant changes in the overproduction factors (compare the green patterns in the two panels of Fig.~\ref{fig:Abundances2})
Full stellar models would be required to get a more accurate estimate of the overproduction factors.

\section{Conclusions}

For the first time, available data on the energies, spins, parities and partial widths of excited states in $^{21}$Ne have been thoroughly evaluated including a careful consideration of their ambiguities and uncertainties. In addition, 
states in $^{21}$Ne have been populated via the $^{20}$Ne($d,p$)$^{21}$Ne reaction, using an implanted $^{20}$Ne target. Angular distributions and neutron widths for states within the Gamow window for massive-star He-core burning were extracted. By combining these data with the evaluated data, reaction rates for $^{17}$O($\alpha,n$)$^{20}$Ne and $^{17}$O($\alpha,\gamma$)$^{21}$Ne have been calculated using updated excitation energies, $J^\pi$ assignments and experimentally derived neutron widths. Using the {\sc RatesMC} Monte Carlo code, uncertainties have been estimated consistently for the first time. Our recommended rates indicate enhanced s-process abundances between Sr and Pb. 
Production of these elements via the enhanced weak s-process in massive stars significantly shortens the timescale for the production of Pb (otherwise only produced via the main s-process in low-mass stars with much longer lifetimes) and provides an alternative to the r-process for producing elements between Fe and Ba in the early Universe.
Experimental constraints on the $\alpha$-widths of the key states in $^{21}$Ne are thus
crucial to allow the production of elements above barium in such massive stars,
and so for the evolution of elements heavier than iron in the early Universe to be understood.

\section*{acknowledgements}

The authors thank Nick Keeley and Andreas Best for useful discussions.  
JFS, AML, CB and CD thank the UK Science and Technology Facilities Council (STFC). PA thanks the trustees and staff of Claude Leon Foundation for support in the form of a Postdoctoral Fellowship. RL, CM, FP, and KS thank the U.S. Department of Energy, Office of Science, Office of Nuclear Physics, for support under Grant No. DE-SC0017799 and Contract No. DE-FG02-97ER41041.
This work was supported by the Fonds de la Recherche Scientifique-FNRS under Grant No IISN 4.4502.19. This paper is based upon work from the \textquoteleft ChETEC\textquoteright\ COST Action (CA16117), supported by COST (European Cooperation in Science and Technology). RH acknowledges support from the IReNA AccelNet Network of Networks, supported by the National Science Foundation under Grant No. OISE-1927130 and from the World Premier International Research Centre Initiative (WPI Initiative), MEXT, Japan. RH also acknowledges funding from the European Union’s Horizon 2020 research and innovation programme under grant agreement No 101008324 (ChETEC-INFRA).

\section*{Data Availability}
The nuclear physics data used in the evaluation are available from NNDC (https://www.nndc.bnl.gov) via ENSDEF and EXFOR. The reaction rates in tabular form will be available as part of the Starlib reaction rate library (https://github.com/Starlib/Rate-Library). Other data arising from the present work are available on reasonable request to the corresponding author.

\newpage
\bibliographystyle{mnras}
\bibliography{references}

\bsp	
\label{lastpage}
\end{document}